\documentclass[11pt]{article}
\newcommand{\Comment}[1]{{}}
\usepackage{amsfonts,amsthm,amsmath,amssymb,slashed}
\usepackage[textwidth = 430 pt, textheight = 630 pt]{geometry}
\usepackage{color} 
\usepackage{indentfirst}

\Comment{\usepackage{color}
\definecolor{MyDarkBlue}{rgb}{0.15,0.15,0.45}
\usepackage[linktocpage=true]{hyperref}
\hypersetup{
colorlinks=true,
citecolor=MyDarkBlue,
linkcolor=MyDarkBlue,
urlcolor=MyDarkBlue,
pdfauthor={Horatiu Nastase and Jacob Sonnenschein},
pdftitle={Euler fluid as a gauge theory, and an action for the Euler fluid},
pdfsubject={hep-th}
}}

\usepackage{hypernat}
\usepackage{graphicx}
\usepackage{cite}

\newcommand\ignore[1]{}
\def\one{{\,\hbox{1\kern-.8mm l}}}

\def\d{\partial}

\newcommand{\Cset}{{\,\,{{{^{_{\pmb{\mid}}}}\kern-.45em{\mathrm C}}}}}

\newcommand{\be}{\begin{equation}}
\newcommand{\bea}{\begin{eqnarray}}

\newcommand{\ee}{\end{equation}}
\newcommand{\eea}{\end{eqnarray}}

\begin{document}

\renewcommand{\thefootnote}{\fnsymbol{footnote}}

\makeatletter
\@addtoreset{equation}{section}
\makeatother
\renewcommand{\theequation}{\thesection.\arabic{equation}}

\rightline{}
\rightline{}
%   \vspace{1.8truecm}

%\begin{flushright}
% preprint nrs.
%\end{flushright}

%\vspace{10pt}

%%%%%%%%%%%%%%%%%

\begin{center}
	{\LARGE \bf{\sc Rotational holographic transport from KN-AdS black hole}}
\end{center}
\vspace{1truecm}
\thispagestyle{empty} \centerline{
	{\large \bf {\sc Pedro Meert${}^{a}$}}\footnote{E-mail address: \Comment{\href{mailto:pedro.meert@unesp.br}}{\tt
			pedro.meert@unesp.br}}
	{\bf{\sc and}}
	{\large \bf {\sc Horatiu Nastase${}^{a}$}}\footnote{E-mail address: \Comment{\href{mailto:horatiu.nastase@unesp.br}}
		{\tt horatiu.nastase@unesp.br}}
}

\vspace{.5cm}

%\vspace{.3cm}

\centerline{{\it ${}^a$Instituto de F\'{i}sica Te\'{o}rica, UNESP-Universidade Estadual Paulista}}
\centerline{{\it R. Dr. Bento T. Ferraz 271, Bl. II, Sao Paulo 01140-070, SP, Brazil}}
%\vspace{.3cm}
%\centerline{{\it ${}^b$School of Physics and Astronomy,}}
%\centerline{{\it The Raymond and Beverly Sackler Faculty of Exact Sciences, }} 
%\centerline{{\it Tel Aviv University, Ramat Aviv 69978, Israel}}
%\vspace{.3cm}
%\centerline{{\it ${}^c$ Simons Center for Geometry and Physics,}} 
%\centerline{{\it SUNY, Stony Brook, NY 11794, USA }} 
\vspace{1truecm}

%%%%%%%%%%%%%%%%%
\thispagestyle{empty}

\centerline{\sc Abstract}

\vspace{.4truecm}

\begin{center}
	\begin{minipage}[c]{380pt}
		{\noindent We consider rotational holographic transport in strongly coupled 2+1 dimensional systems, from the point of
			view of 3+1 dimensional gravity. We consider the moment of inertia $I$ as a kind of transport coefficient, identified
			with the moment of inertia of a charged rotating black hole in $AdS_4$ background. In the low-temperature region, we
			find the behaviour of the density $I/A$ with temperature $T$ and angular 	velocity $\Omega$, and find a quadratic
			behaviour for $\d(I/A)/\d\Omega$ with $T$, in the presence of some charge $Q$.	}
	\end{minipage}
\end{center}

\vspace{.5cm}

\setcounter{page}{0}
\setcounter{tocdepth}{2}

\newpage

\tableofcontents
\renewcommand{\thefootnote}{\arabic{footnote}}
\setcounter{footnote}{0}

\linespread{1.1}
\parskip 4pt

\section{Introduction}\label{Sec:Intro}

The AdS/CFT correspondence (see the book~\cite{Nastase:2015wjb} for a review)
was initially a top-down construction in string theory, starting with Maldacena's
seminal paper~\cite{Maldacena:1997re}. In the case of specific theories, related to $AdS_n\times X_m$ backgrounds,
for instance, ${\cal N}=4$ SYM vs. $AdS_5\times S^5$, one has a map between perturbative classical gravity
and nonperturbative field theory, or vice versa, starting with~\cite{Witten:1998qj,Gubser:1998bc}.
Finite temperature for the field theory was introduced in~\cite{Witten:1998zw} by having black holes in the AdS background,
which could, in those cases, be thought of as a near-horizon near-extremal limit of non-extremal branes~\cite{Itzhaki:1998dd}.

The phenomenological extension of the original AdS/CFT correspondence led both to non-AdS backgrounds (the ``gauge/gravity
duality''), and to general theories in AdS background, not necessarily coming from string theory, for which one of the
most common applications are to strongly coupled condensed matter systems (``AdS/CMT'', see the book~\cite{Nastase:2017cxp}
for a review). Since these are usually at finite temperatures, black holes in AdS backgrounds are usually used for the description
of more realistic systems.

One of the most relevant phenomena in condensed matter is transport, so holographic transport has been the subject of
a lot of research (see, for instance,~\cite{Hartnoll:2009sz,Nastase:2017cxp}). Usually, one focuses on the transport of electric
charge and heat, with the transport coefficients being electric and thermal conductivities, and thermo-electrical (mixed) ones.

But it is interesting to consider also rotation in the field theory for the condensed matter system, which would correspond
to considering a rotating black hole in AdS, so Kerr-Newman-AdS black hole. This would be, for instance, the case of
a Bose condensate at very low temperatures (for instance, after using evaporative cooling to get to super-low temperatures),
rotating in a magnetic trap. In that case, it would be very hard to measure electric
or heat transport, but rotation can be thought of as a different kind of transport. Indeed, we can impose rotation at a certain
angular velocity $\omega$ (input), for instance by fixing $\omega$ for the trap and find the resulting angular momentum $J$ of the
condensed matter material (response of the system). {\em In linear response theory}, we would have
\be
J=I \omega\;,
\ee
where $I$, the momentum of inertia of the material, would be defined as the transport coefficient in this case.
Then, this must be understood as $\delta J=I\delta \omega$, for $\delta \omega\rightarrow 0$.

Rotating field theory systems via holography
have been considered before in the context of strongly coupled Quark-Gluon Plasma (sQGP), for example
in~\cite{McInnes:2014haa,Braga:2022yfe}, while in the context of condensed matter, only in the context of
holographic superconductors, see for instance \cite{Sonner:2009fk}.

However, we would like to understand them a bit more generally, in the context of more general thermodynamics,
specifically black hole thermodynamics on the gravitational side, as well as to study the implications of rotation viewed
as transport.

The most general axially symmetric black hole solution to Einstein's equations is the  Kerr-Newman one
containing mass, charge, and angular momentum. This solution also exists on a background with a negative cosmological
constant and, therefore, can be used in AdS/CFT, as its asymptotic behavior
is well understood.

In~\cite{Hawking:1998kw}, shortly after the AdS/CFT correspondence was defined, the effects of
rotation were investigated under the duality. In particular, the angular velocity at the boundary of
space-time was shown to be finite, and therefore the relation to the field theory
has to be appropriately defined, such that the thermodynamics
of the black hole is described correctly. The finite angular velocity at the boundary translates to rotation
in the dual field theory, with a modification that we will present.

The paper is organized as follows. In Section~\ref{Sec:KN-AdS} we briefly review the most important properties
of the Kerr-Newman-AdS space-time that will be relevant to our subsequent description of thermodynamics and
moment of inertia, presenting the low-temperature expansion explicitly. Section~\ref{sec:Results} contains the
main results for the angular velocity and moment of inertia of this solution. We also present an
interpretation of the dual field theory and how the moment of inertia can be seen as a response function in
the thermodynamic sense. Finally, in Section~\ref{Sec:Conclusion} we summarize the results and discuss open
possibilities.
% Changed wording in this paragraph.

\section{Charged rotating black hole in AdS}\label{Sec:KN-AdS}

The Kerr-Newman solution of Einstein's equations describes an axially
symmetric black hole with mass, angular momentum and electric charge. It was first
introduced in~\cite{Newman:KN-sol1965},
in an asymptotically flat background. The same type of solution was then obtained in an
asymptotically AdS space-time background~\cite{Carter:1968ks},
and is particularly interesting in the context of holographic duality, as the
theory on the boundary has an associated angular velocity. In Boyer-Lindquist
coordinates the line element reads
\begin{equation}
	\label{Eq:BLcoord-KNAdS}
	ds^{2}=
	-\frac{\Delta_{r}}{\rho^{2}}
	\left[
		dt-\frac{a}{\Xi}\sin^{2}\theta d\phi
		\right]^{2}
	+\frac{\rho^{2}}{\Delta_{r}}dr^{2}
	+\frac{\rho^{2}}{\Delta_{\theta}}d\theta^{2}
	+\frac{\sin^{2}\theta\Delta_{\theta}}{\rho^{2}}
	\left[
		adt-\frac{\left(r^{2}+a^{2}\right)}{\Xi}d\phi
		\right]^{2},
\end{equation}
where we have defined
\begin{align}
	\begin{aligned} \label{Eq:metric-fncts}
		\rho^{2}        & =r^{2}+a^{2}\cos^{2}\theta,                                    \\
		\Delta_{r}      & =\left(r^{2}+a^{2}\right)\left(1+l^{-2}r^{2}\right)-2mr+Q^{2}, \\
		\Delta_{\theta} & =1-l^{-2}a^{2}\cos^{2}\theta,                                  \\
		\Xi             & =1-l^{-2}a^{2},
	\end{aligned}
\end{align}
and use units such that $ c = G_N = 1 $. The parameter $ a $ is associated with
the rotation of the black hole, $ m $ to its mass, $ Q $ to the electric
charge, and
\begin{equation}\label{Eq:AdSrad}
	l^{-2} = - \frac{\Lambda}{3}
	\; ,
\end{equation}
is the AdS radius, and is defined relative to the negative
cosmological constant $ \Lambda $.

As discussed in~\cite{Hawking:1998kw,Caldarelli:1999xj},
the solution is regular as long as some conditions are satisfied. These conditions can be seen as constraints
on the parameters $ m, a, Q $, and are obtained from investigation of the event horizon, which is defined by
\begin{equation} \label{Eq:delta0}
	\Delta _{r _{+}} = 0 \ ,
\end{equation}
$ r _{+} $ being the largest root of the equation. It can be shown that the
cosmic censorship is satisfied only if the mass parameter is above a certain
critical value,
\begin{align} \label{Eq:M_critical}
	m_{c}=
	\frac{l}{3\sqrt{6}}
	 & \left[
	\sqrt{\left(1+l^{-2}a^{2}\right)^{2}+12l^{-2}\left(a^{2}+Q^{2}\right)}
	+2\left(1+l^{-2}a^{2}\right)
	\right]\times \nonumber \\
	 & \left[
	\sqrt{\left(1+l^{-2}a^{2}\right)^{2}+12l^{-2}\left(a^{2}+Q^{2}\right)}
	-\left(1+l^{-2}a^{2}\right)
	\right]^{1/2}
	.
\end{align}

The parameter $ a $ is also bounded by $ a^{2} \leq l^{2} $. When the inequality is saturated, the
boundary of the space-time rotates at the speed of light. This defines an extremal black hole and was
thoroughly investigated in~\cite{Hawking:1998kw}.
% Changed wording in this paragraph.

\subsection{Thermodynamic variables}\label{Subsec:Thermodynamics}

We quote the main results for the Kerr-Newman-AdS thermodynamics, which was
extensively studied in~\cite{Caldarelli:1999xj}.
The temperature is obtained from the regularity of the solution with analytical continuation of the time coordinate
$ t\to i\tau $ and rotation
parameter $ a\to ia $. The regularity conditions at the horizon require the periodicity conditions
$ \tau\sim\tau+\beta $ in $ \tau $, and $ \phi\sim\phi+i\beta\omega\left(r\to r_{+}\right) $ in the polar
angle.
Explicitly, the temperature is given by
\begin{equation} \label{Eq:T-1}
	\beta \equiv
	T^{-1} =
	\frac{4\pi\left(r_{+}^{2}+a^{2}\right)}
	{r_{+}
		\left(
		1+a^{2}l^{-2}+3r_{+}^{2}l^{-2}-\left(a^{2}+Q^{2}\right)r_{+}^{-2}
		\right)}
	.
\end{equation}

The area of the (outer) horizon $ r_+ $ can be computed directly from the metric, giving
\begin{equation} \label{Eq:Hor_area}
	A=\frac{4\pi\left(r_{+}^{2}+a^{2}\right)}{\Xi}\ ,
\end{equation}
where $ \Xi $ was defined in Eq.~\eqref{Eq:metric-fncts}. The area in the
equation above is related to the entropy by the Bekenstein-Hawking formula
$ S = \frac{A}{4} $ (for $G_N=1$).
% TODO: Add ref here for Bekenstein?

The angular velocity and angular momentum are the most relevant quantities for
our work. Space-times with non-vanishing angular momentum in asymptotically AdS
backgrounds are particularly interesting because the angular velocity at the
boundary is finite~\cite{Hawking:1998kw}.
For general $ r $ the angular velocity reads
\begin{equation} \label{Eq:omega-general}
	\omega=
	\frac{a\Xi}{\Sigma^{2}}
	\left[
		\Delta_{\theta}\left(r^{2}+a^{2}\right)-\Delta_{r}
		\right]
	.
\end{equation}

This can be obtained by writing Eq.~\eqref{Eq:BLcoord-KNAdS} in the so-called normal
form,
\begin{equation}
	ds^{2} =
	- N^{2}dt^{2}
	+ \frac{\rho^{2}}{\Delta_r}dr^{2}
	+ \frac{\rho^{2}}{\Delta_{\theta}}d\theta^{2}
	+ \frac{\Sigma^{2}\sin^{2}\theta}{\rho^{2}\Xi^{2}}
	\left(
	d\phi - \omega dt
	\right)^{2}\;,
\end{equation}
where
\begin{equation}
	\Sigma^{2} =
	\left(r^{2}+a^{2}\right)^{2}\Delta _{\theta}
	- a^{2} \Delta _{r} \sin^{2}\theta
	\; ,
\end{equation}
and
\begin{equation}
	N^{2}=
	\frac{\rho^{2}\Delta_r\Delta _{\theta}}{\Sigma^{2}}
	\ .
\end{equation}

The ``thermodynamic angular velocity'' is given by the difference between the value of the angular velocity at the horizon
and at infinity,
\begin{equation} \label{Eq:omega-thermo}
	\Omega=
	\lim_{r\to r_{+}}\omega-
	\lim_{r\to\infty}\omega=
	\frac{a\left(1+r_{+}^{2}l^{-2}\right)}{r_{+}^{2}+a^{2}} .
\end{equation}

Note that this is different from the result in flat space, where there is no rotation of the solution at $r\rightarrow \infty$.
In AdS space, because of the finite value at infinity, the relation to the boundary field theory is given by the above difference.

The angular momentum and total mass can be computed by
Komar integral methods, whereas the total electric charge is obtained by using
the computation of electric potential flux, as in~\cite{Papadimitriou:2005ii}.
One finds, in terms of the constant parameters $a,m,Q$ of the solution, that the total angular momentum, total mass and total
electric charge are given by
\begin{equation} \label{Eq:net_J}
	J = \frac{am}{\Xi^{2}}
	\ ,
\end{equation}
\begin{equation} \label{Eq:net_M}
	M = \frac{m}{\Xi^{2}}
	\ ,
\end{equation}
\begin{equation} \label{Eq:net_Q}
	\mathcal{Q} = \frac{Q}{\Xi}
	\ .
\end{equation}

Finally, the electric potential is obtained similarly to the angular velocity:
one computes the difference between its value at infinity and the horizon,
\begin{equation} \label{Eq:e_potential}
	\Phi=
	\frac{Q r _{+}}{r _{+}^{2} + a^{2}}
	\ .
\end{equation}

These are all the thermodynamic variables of the system, which can be used to write
the relevant thermodynamic potentials.

\subsection{Euclidean action and thermodynamic potentials}
\label{Subsec:EA-Thermo}

The metric~\eqref{Eq:BLcoord-KNAdS} is a solution of the equations of motion
obtained from the action
\begin{equation} \label{Eq:EHM-action}
	S=-\frac{1}{16\pi}\int d^{d}x\sqrt{-g}\left(R-2l^{-2}-F^{2}\right)
	\; ,
\end{equation}
with $ F _{ab}=2\partial _{[a}A _{b]} $ the field strength of the
electromagnetic field, and $ F^{2}=F _{ab}F^{ab} $. The gauge potential reads
\footnote{
	The most general form of the potential contains another term dependent on the
	magnetic charge. Here we're setting it to zero from the beginning.
}
\begin{equation} \label{Eq:Maxwell_potential}
	A = -\frac{Qr}{\rho^{3}}
	\left(
	dt - \frac{a}{\Xi^{2}} \sin^{2}\theta d\phi
	\right)
	\ .
\end{equation}
%
%TODO: Review this paragraph

To obtain a finite result for the action (which is mapped to the thermodynamics potential), we have to do
the holographic renormalization~\cite{Henningson:1998gx,Balasubramanian:1999re,deHaro:2000vlm,Bianchi:2001kw}.
This process involves adding 2 terms, one associated with the boundary and a
counter-term necessary to remove divergences.
These terms won't affect the equations of motion and are included with the only
purpose of making the (Euclidean) on-shell action finite. Explicitly, the expression is
\begin{equation} \label{Eq:Complete_action}
	I=I_{\text{bulk}}+I_{\partial}+I_{\text{C-T}}=
	-\frac{1}{16\pi G}\int_{\mathcal{M}}d^{4}x\sqrt{g}\left[R-2\Lambda-F^{2}\right]
	-\frac{1}{8\pi G}\int_{\partial\mathcal{M}}d^{3}x\sqrt{h}K
	+I_{\text{C-T}}
	\ ,
\end{equation}
where we have already written $ t \mapsto it $. The second term is the Gibbons-Hawking term,
a surface term depending on the induced metric at
the boundary, $ h _{ij} $, and the trace of the extrinsic curvature $ K =  \nabla_{j}n^{j} $, where $ n^{j} $
is the vector orthonormal to $ h_{ij} $.

The counter-term, $ I _{\text{C-T}} $ has a standard form~\cite{Emparan:1999pm}\footnote{Potential curvature 
squared terms, ${\cal R}_{ij}{\cal R}^{ij}$ and ${\cal R}^2$, vanish on the boundary at infinity, so were omitted.}
\begin{equation} \label{Eq:Action_CT}
	I_{\text{C-T}}=
	\int_{\partial\mathcal{M}}d^{3}x
	\sqrt{h}
	\left[
		\frac{2}{l} +\frac{l\mathcal{R}}{2}
		%-\frac{l^{3}}{2}\left(\mathcal{R}_{ij}\mathcal{R}^{ij}-\frac{3}{8}\mathcal{R}^{2}\right)
		\right]
	\ ,
\end{equation}
where $ \mathcal{R} $ is the Ricci scalar, of the induced metric at
the boundary.

Choosing the boundary metric to be $ \partial\mathcal{M}=S^{1}\times S^{2} $,
where $ S^{1} $ is the time circle, and $ S^{2} $ a 2-sphere with large radius,
which is sent to infinity after integration, the on-shell action is
\begin{equation} \label{Eq:Onshell-action}
	I=
	\frac{\beta}{4l^{2}\Xi}
	\left[
	-r_{+}^{3}
	+l^{2}\Xi r_{+}
	+\frac{l^{2}\left( a^{2} + Q^{2} \right)}{r_{+}}
	+\frac{2l^{2}Q^{2}r_{+}}{\left(r_{+}^{2}+a^{2}\right)}
	\right]
	.
\end{equation}

The on-shell action~\eqref{Eq:Onshell-action} is related to the thermodynamic
potential via the temperature,
\begin{equation}\label{Eq:G_FreeEnergy}
	G\left(T,\Omega,\Phi\right)=TI
	\ ,
\end{equation}
from which we can, in principle, extract the other relevant quantities by taking
derivatives, such as the entropy $ S $, the angular momentum $ J $ and the electric
charge $ Q $. However, this is a very hard task to do analytically, due to
the relation between the black hole parameters $ r_+, a, Q $, and the
thermodynamic quantities presented in the Section~\ref{Subsec:Thermodynamics}.

\subsection{Low temperature expansion}\label{sec:low_temperature_expansion}

To completely rewrite the thermodynamics in terms of the variables presented in subsection~\ref{Subsec:Thermodynamics}, we
need to solve a system of 3 equations relating $ \{r_+,a,Q\} $ to $ \{T,\Omega,\mathcal{Q}\} $. These  are
Eqs.~\eqref{Eq:T-1},~\eqref{Eq:omega-thermo} and~\eqref{Eq:net_Q}. This is a non-linear system of equations,
and we resort to approximations to solve for the temperature, which is the variable we are most interested in,
and to numerical methods to solve for the rotation parameter $ a $. The relation between $ Q $ and $ \mathcal
	Q$ is straightforward from Eq.~\eqref{Eq:net_Q}.

Eq.~\eqref{Eq:T-1} has 4 distinct roots when we solve it for $ r_+\left( T \right) $. To choose one of them, we
check which are real and greater than zero for any $ a,Q $. The only root that satisfies these criteria is
\begin{equation}\label{Eq:r0_T0}
	r _{+} = r_{0} =
	\frac{\sqrt{-a^{2}-l^{2}+\eta}}{\sqrt{6}}
	,
\end{equation}
where
\begin{equation}\label{Eq:eta}
	\eta = \sqrt{a^{4} + l^{4} +14a^{2}l^{2} + 12Q^{2}l^{2}}.
\end{equation}

From the definition of $ r _{+}$, Eq.~\eqref{Eq:delta0}, we have
\begin{equation}\label{Eq:m_t0}
	m =
	\frac{1}{2r_{+}}
	\left[
	\left(r_{+}^{2}+a^{2}\right)\left(1+l^{-2}r_{+}^{2}\right)+Q^{2}
	\right].
\end{equation}
For $ r _{0}$ given by~\eqref{Eq:r0_T0}, one finds that $m=m_{c}$, meaning that we are close to the
extremality regime of the black hole.

We rewrite the temperature as
\begin{equation}
	T\left( r_0 + \delta r \right) \approx A\left( a,Q \right) \delta r ,
\end{equation}
where we keep $ \delta r $ up to the first order in the expansion. We find the function $ A\left( a,Q \right) $ is
\begin{equation}
	A\left( a,Q \right) =
	\frac{3\left[ \eta^{2} - \eta\left( a^{2} + l^{2} \right) \right]}
	{l^{2}\pi\left( l^{2} - 5a^{2} - \eta \right)\left( a^{2} + l^{2} - \eta \right)} .
\end{equation}
Now if we let $ T $ become an independent variable we can rewrite $ \delta r = TA^{-1}\left( a,Q \right) $,
and have $ T\ll 1 $, such that the condition for $ \delta r \ll 1 $ is always satisfied. So by replacing
\begin{equation}\label{Eq:pertR}
	r_+ \mapsto r_0 + \delta r ,
\end{equation}
in the expressions~\eqref{Eq:omega-thermo} and~\eqref{Eq:net_J}, for angular velocity and angular momentum,
we have them close to zero temperature.

\section{Moment of inertia and holographic rotational transport}\label{sec:Results}

\subsection{Moment of Inertia}\label{Subsec:MomIn}

We are interested in the thermodynamics and holographic transport of the dual system, based on the formulas from the
previous section. As mentioned in the introduction, the relevant quantity is the moment of inertia of the system, defined by
\begin{equation}\label{Eq:MI_def}
	I = J/\Omega\; ,
\end{equation}
where $ \Omega $ and $ J $ were defined in Eqs.~\eqref{Eq:omega-thermo} and~\eqref{Eq:net_J}, respectively.
To obtain the expression in terms of the temperature we note that the angular momenta~\eqref{Eq:net_J} can be
written as
\begin{equation}\label{Eq:J_m}
	J = \frac{a}{2\Xi^{2}r_+}
	\left[
		\left(r^{2}_{+}+a^{2}\right)\left(1+l^{2}r^{2}_{+}\right) + q^{2}
		\right],
\end{equation}
where the definition of $ r_+ $ was used to write $ m $ in terms of the horizon radius. The low-temperature
expansion is achieved by the replacement $ r_{+}\to r_0 + \delta r $, as in~\eqref{Eq:pertR}.

To avoid considerations regarding the physical size of the dual object being described by the dual theory, we
will investigate the behavior of the density of the moment of inertia, $I/A$. The area $ A $ is given by
Eq.~\eqref{Eq:Hor_area}, and $ I = J/\Omega $ is first determined in terms of parameters $ \{ r_+,a,Q\} $,
then solved for $ r_+ $ and $ Q $ in terms of $ T,\mathcal{Q} $. Finally, we will use numerical simulations to plot these
as functions of the rotation parameter $ a $ and solve the equation in terms of $ \Omega $.

From now on we consider $ l=10 $, and therefore restricting $ a\leq l $. Choosing this value for $ l $
ensures that the cosmological constant $ \Lambda\sim 10^{-2}\ll 1 $, from the definition of the AdS radius in
Eq.~\eqref{Eq:AdSrad}. The action then is just the Einstein-Hilbert
with a Maxwell term, allowing us to neglect stringy corrections and apply the classical gravity approximation.

We are interested in results that are expressed in terms of physical variables, obeying physical constraints, which can also
be interpreted from the point of view of the dual field theory. The physical variables are $(J,M,{\cal Q})$ and (angular momentum,
mass, charge) and not the black hole parameters $(a,m,Q)$, as well as $\Omega, A$ (angular velocity, area).

The angular velocity must be smaller than the speed of light, and this becomes a constraint on the relevant physics.
%so we compute it in terms of the rotation parameter $ a $ for different values of charge. 
As discussed in~\cite{Caldarelli:1999xj}, the angular velocity
will be smaller than the speed of light when $ \Omega < 1/l $, in which case one can define a time-like
Killing vector globally outside the event horizon, such that the black hole is in thermal equilibrium with
thermal radiation all the way to infinity, and thermodynamics can be consistently defined.

\begin{figure}[ht!]
	\centering
	\includegraphics[scale=0.6]{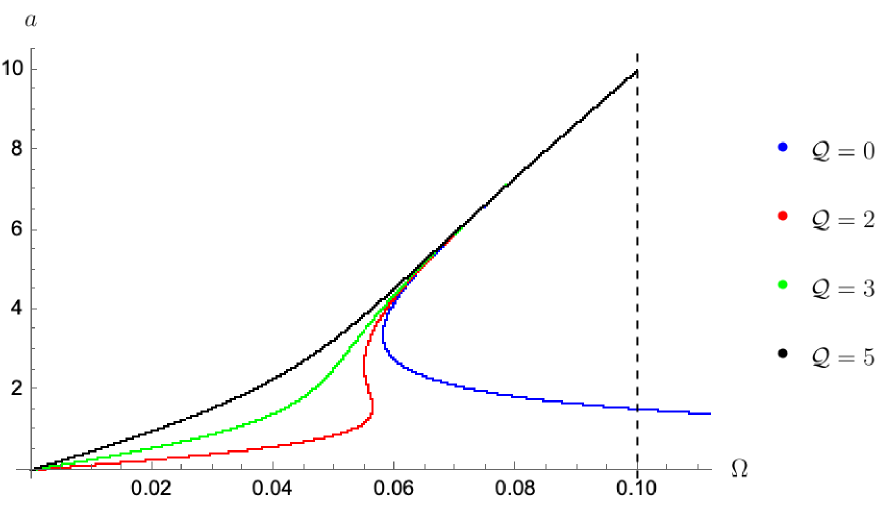}
	\caption{Angular velocity $\Omega$ and corresponding values of $ a $. The temperature is fixed at $ T=0.1 $.}
	\label{fig:Wxa_q}
\end{figure}

%The angular velocity for each value of $ a $ is displayed 
Even though $a$ is not a physical parameter, we consider its variation with the physical velocity $\Omega$, as a function of the
charge ${\cal Q}$ (for fixed $T$, understood as the temperature of the dual field theory),
in Fig.~\ref{fig:Wxa_q}. We observe that at physical charge ${\cal Q}=0$ (and thus also at parameter $ Q=0$)
for small $a$, we are in the
unphysical region of $\Omega>1/l=0.1$. For a clearer picture, we plot $a$ vs. $\Omega$ for ${\cal Q}=0$ and ${\cal Q}=3$, for
$T=0$ and $T=0.1$, finding that ${\cal Q}=0$ is unphysical for small $a$ (small $J$).
From the black hole perspective, it is clear that small temperatures are achieved when we are near criticality, so small ${\cal Q}$ means
that $a$ must be close to its maximal value, and then $\Omega$ is also large. But from the point of physical parameters (interpretable,
and tunable in the dual field theory), it is less clear.

\noindent

\noindent
\begin{figure}[ht!]
	\centering
	\includegraphics[scale=0.4]{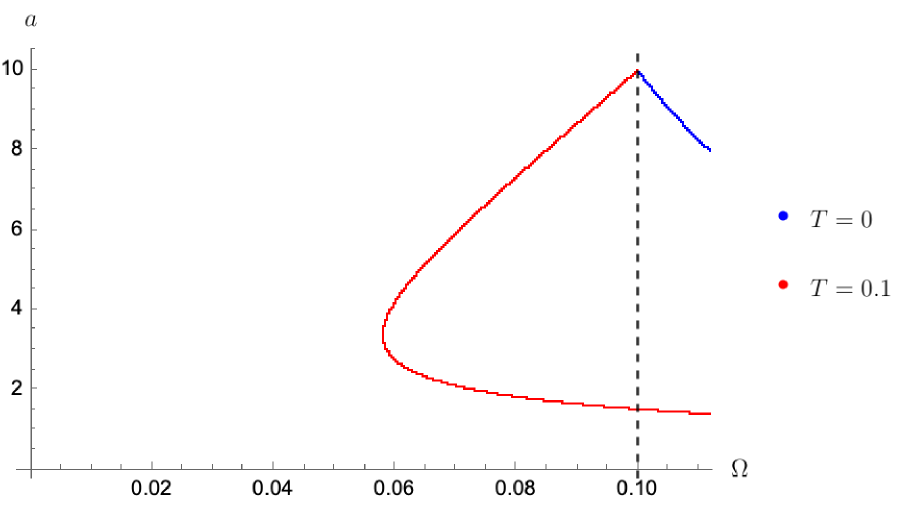}
	\includegraphics[scale=0.4]{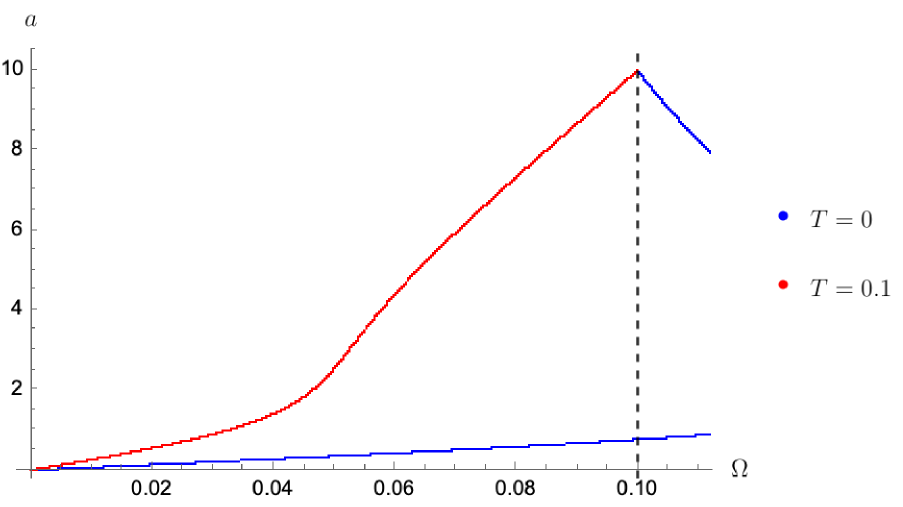}
	\caption{$a(\Omega)$ for $T=0$ and $T=0.1$. Left: Charge  $ {\cal Q} = 0 $. Right: Charge  $ {\cal Q} = 3 $.}
	\label{fig:Wxa_T}
\end{figure}

% Isn't this paragraph duplicated, considering the same consideration has been mentioned above?

The moment of inertia grows with system size, and the relevant system size in the black hole is its area, mapped to the
area (``volume'') in the 2+1 dimensional field theory. 
Then, as we already mentioned, if we want to obtain results that are independent on
the area, we should calculate densities, in particular the density $I/A$, moment of inertia over area.

\begin{figure}[ht!]
	\centering
	\includegraphics[scale=0.5]{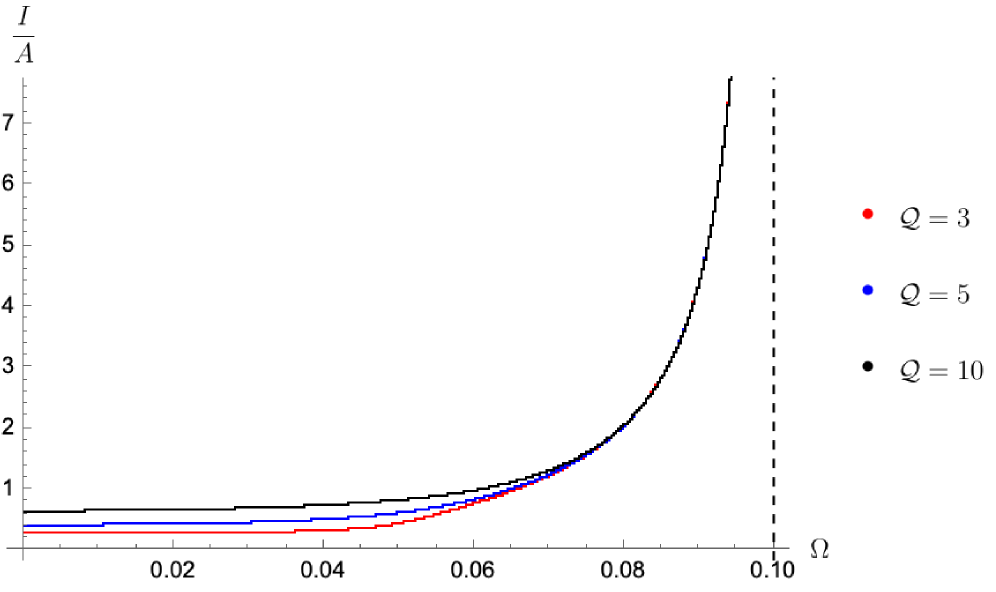}
	\caption{Moment of inertia over area, $I/A$, as a function of angular velocity $\Omega$, for nonzero
		charge. The temperature is fixed at $ T=0.1 $.}\label{fig:IAxW_Q}
\end{figure}

\begin{figure}[ht!]
	\centering
	\includegraphics[scale=0.4]{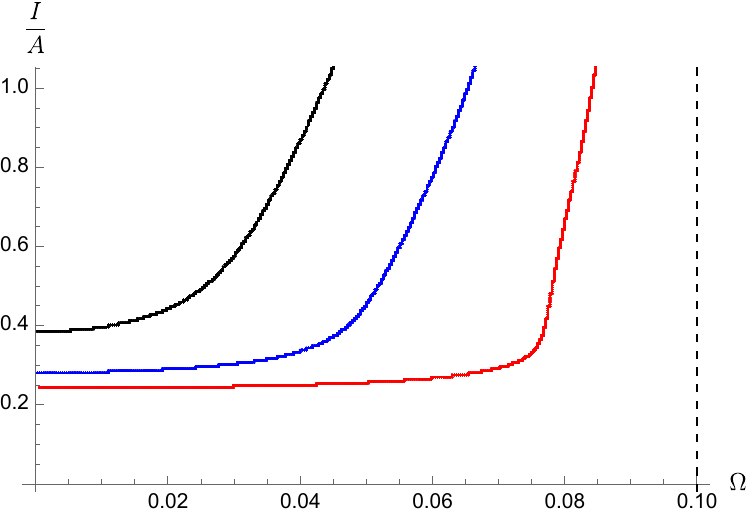}
	\includegraphics[scale=0.4]{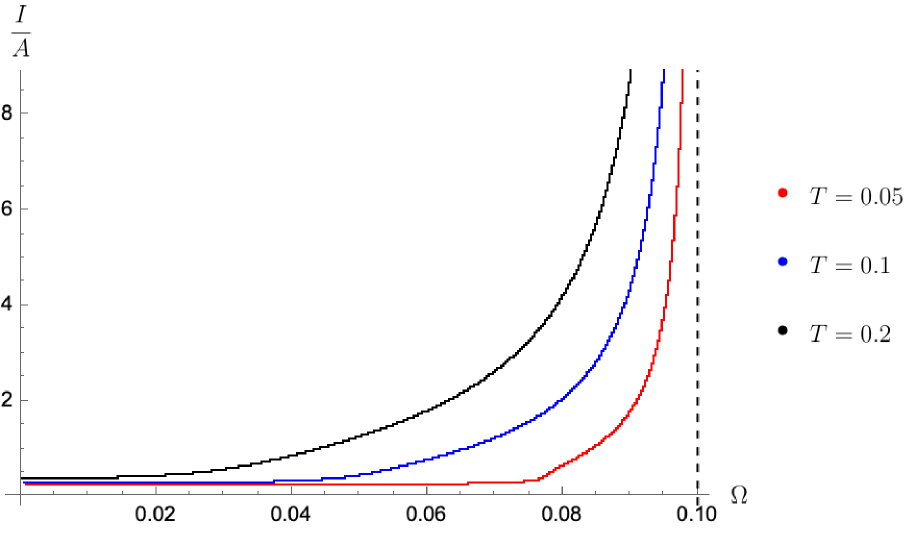}
	\caption{Moment of inertia over area, $I/A$, vs. angular velocity $\Omega$, for different temperatures. The charge
		is fixed at $ {\cal Q}=3. $}\label{fig:IAxW_T}
\end{figure}

We first plot the density of the moment of inertia $I/A$ as a function of $\Omega$, at various fixed ${\cal Q}$'s, but for $T=0.1$,
in Fig.~\ref{fig:IAxW_Q}.
At fixed temperature and different (fixed) electric charges, the main difference with higher charges is the value
of $ \frac{I}{A} $ at zero angular velocity.
Then in Fig.~\ref{fig:IAxW_T} we plot the density of the moment of inertia $ I/A $ as a function of $\Omega$ at fixed ${\cal Q}=3$, but
for different temperatures.
Both pictures
correspond to the same plot, the difference is the left panel zooms in for smaller values of $I/A$.

\noindent

\noindent

\noindent

Values of $\Omega$ that are interesting from the point of view of the dual field theory are ones that are well below the speed of
light (the maximum line at $\Omega=0.1$). But we see that in these cases, $I/A$ is almost constant as a function of $\Omega$,
the only change is that $I/A(\Omega=0)$ depends on $T$ and ${\cal Q}$. This is reasonable since the moment of
inertia should not go to zero at $T=0$ or $\Omega=0$, but rather to a constant.

\begin{figure}[ht!]
	\centering
	\includegraphics[scale=0.5]{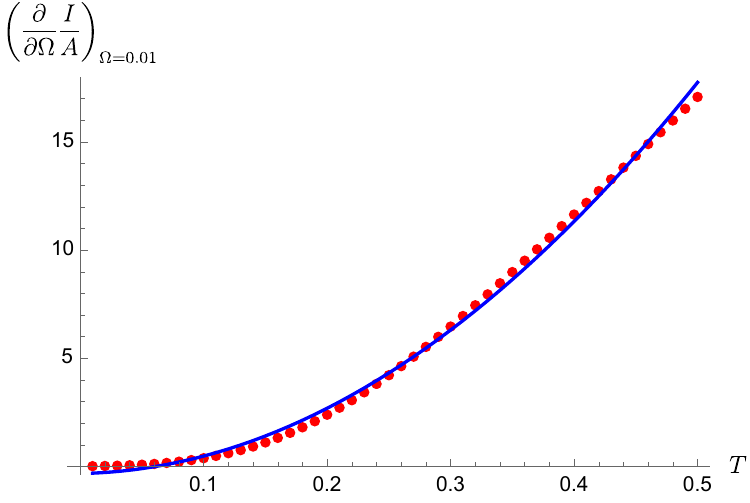}
	\includegraphics[scale=0.5]{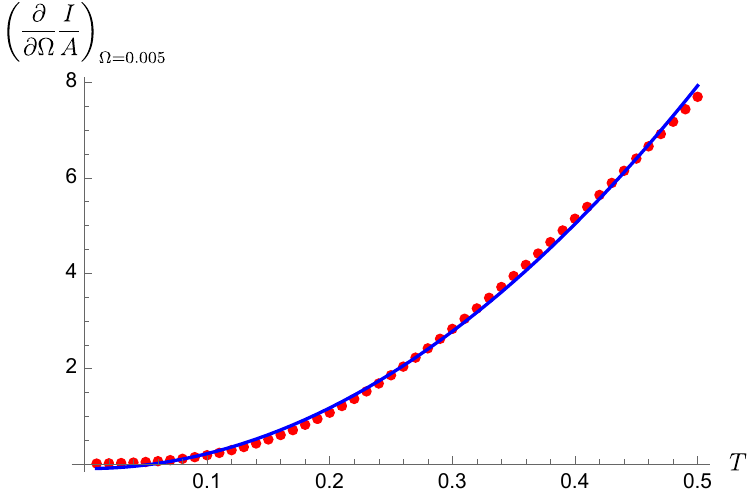}
	\includegraphics[scale=0.5]{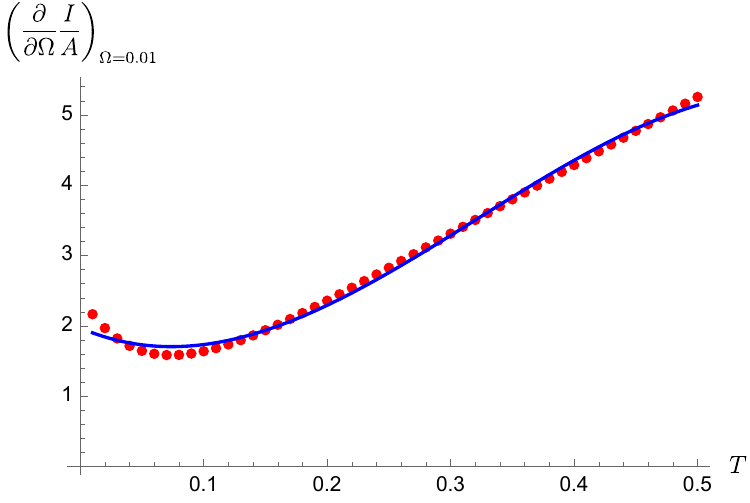}
	\caption{The variation $d(I/A)/d\Omega$ against $T$, at fixed $\Omega$ and ${\cal Q}=3$, for $\Omega=0.001$ (upper left),
		$\Omega=0.0005$ (upper right), and also for ${\cal Q}=100$ and $\Omega=0.001$ (down). }\label{fig:didwxT}
\end{figure}

On the other hand, that means that it is better to plot its variation, $\d(I/A)/\d\Omega$ as a function of $\Omega$ or $T$.
For the purposes of the dual field theory, usually the dependence on the temperature is more interesting.
Therefore, in Fig.~\ref{fig:didwxT} (up), we have plotted $\d(I/A)/\d\Omega$ as a function of $T$, for two (small) values of $\Omega$
and ${\cal Q}=3$. In these cases, the dots are fitted with a quadratic ($ax^2 +bx+c$), as the blue line.

On the other hand, in Fig.~\ref{fig:didwxT} (down) we also plotted the same at much larger ${\cal Q}$, ${\cal Q}=100$,
and we obtain that there is a minimum around
$T=0.2$. This is indicative of a phase transition, for this large ${\cal Q}$ value, though the large value is probably unphysical.

\noindent
%\begin{figure}[ht!]
%	\centering
%	\includegraphics[scale=0.5]{.pdf}
%	\caption{$d(I/A)/d\Omega$ against $\Omega$ for fixed $T$. }\label{fig:IAxW_Q}
%\end{figure}

\noindent

\subsection{Dual theory interpretation}\label{Subsec:DualTh}

It was already found in~\cite{Sonner:2009fk} that the Kerr-Newman-AdS spacetime describes some properties
of rotating superconductors in the strongly coupled regime.

But, more generally, we want to consider the thermodynamics of strongly coupled condensed matter systems
under rotation as coming from the thermodynamics of the rotating black hole in AdS.
As we have seen, one important ingredient, in the presence of rotation, is the overall charge.

In this case, we want to think of rotation as a kind of transport, as explained in the introduction:
we input the angular velocity $\Omega$ (which is a tunable quantity), and in linear response
theory, we find the angular momentum $J=I\Omega$, so the transport coefficient is $I=J/\Omega$.

But on the gravity side, the area (``volume'' in 2+1 dimensions) of the material is mapped to the horizon size $A$ so, in
order to have a result independent of the size of the material, we should divide by it, which is why we have plotted
the density of moment of inertia $I/A$, instead of just $I$.

We have obtained, not surprisingly, since
$I$ characterizes the material just like the effective mass $m_*=p/v$, that $I$ is approximately constant as function of $\Omega$
and $T$ at small temperatures $T$, which are relevant for strongly coupled materials.
We have therefore plotted the derivative $\d(I/A)/\d\Omega$ as a function of $T$, and
we have obtained an approximately quadratic behavior, which should be something that could be experimentally measured.

We have also found that the system needs an overall charge in order to have a sensible map to the gravity dual
at low temperatures and small rotations. This means that the experimental set-up relevant for this cannot be a pure Bose condensate
at very low temperatures, but it must have some electrically charged component, perhaps some free electrons.

If the charge becomes too large, we have found that $\d(I/A)/\d\Omega$ decreases as a function of temperature, and then
starts to increase after a minimum, which
could indicate a phase transition. Except that values of ${\cal Q}$ are large enough so that it probably would break up the
strongly coupled material, and the temperature is large enough that the (small $T$) approximations used are not valid. So we
don't expect this behavior to be relevant in the real world.

\section{Conclusions}\label{Sec:Conclusion}

In this work we have considered holographic rotational transport in 2+1 dimensions, specifically the moment of inertia $I=J/\Omega$,
understood as a transport coefficient, from the point of view of the gravity dual Kerr-Newman-AdS black hole in 3+1
dimensions, and its thermodynamics. Since the expressions we obtained were considerably complicated, we used an
approximation, restricting to low temperatures, which is the domain of most interest for strongly coupled 2+1 dimensional
condensed matter systems. For the relevant formulas, specifically for the density of the moment of inertia $I/A$,
we used numerical approximations and plotted the solutions in Section~\ref{sec:Results}.

We have found that the density of the moment of inertia $I/A$ changes little with $\Omega$ and $T$ for small $T$,
as expected for a quantity that in solids would be a constant, but we have found that its derivative, $\d(I/A)/\d\Omega$ as
a function of $T$ is a rising quadratic form, for not too large a charge ${\cal Q}$, something that could be checked experimentally.
However, we note that having a nonzero charge, which was necessary for a consistent map to the gravity dual, would
mean perhaps that the strongly coupled system (perhaps Bose condensate) would need to have an electron component.

We should also note that the near-extremality of the black hole was a feature of the regime we were interested in the 
field theory: it was necessary in order to have a small temperature $T$, yet a nontrivial rotation $\Omega$ and mass $M$
(as well as a nonzero entropy). Going further away from extremality would imply a larger temperature $T$, but that is 
of less interest in condensed matter (as well as being harder to analyze in gravity).
Also, in this work we have considered a spherical black hole horizon. 
One could consider also a planar horizon, but it is unclear how 
to analyze that in the case of a 2-dimensional rotation in the planar horizon. Other possibilities that have been considered 
are cylindrical or toroidal horizons (since in 2 spatial dimensions, these correspond to just identifications of the plane), 
the former for instance in \cite{Braga:2022yfe}, with cylinder rotation around its axis, 
but it is not clear how they are related to the rotation of a condensed matter system.

For future research,
it would be interesting to study linear transport of velocity, meaning the effective mass $m_*$, relating an input
velocity $v$ to an output momentum $p$, $p=m_*v$, but that would be both harder to simulate in the gravity dual,
and harder to devise an experimental setup.

It would also be interesting to consider the interplay of rotation with other factors, like a magnetic field, which is
present in the relevant cases: the evaporative cooling of Bose-Einstein condensates down to very low temperatures
is usually done in the presence of magnetic traps. The (condensed matter) theory of rotating Bose-Einstein condensate in traps was 
already started, for instance, in \cite{PhysRevB.77.155317}.

%\newpage
\section*{Acknowledgments}
%%%%%%%%%%%%%%%%%%%%%%%%%%%%%%%%%%%%%%%%%%%%%%%%%%%%%%%%%%%%%%%%%%%%%%%%%%%%%%%%%%%%%%%%

We would like to thank Oleg Berman and Caio de Souza for useful discussions at the early stages of this
project, and Pablo de Castro for helpful insights on the implementation of numerical methods used in
Sec.~\ref{sec:Results}.
The work of HN is supported in part by  CNPq grant 301491/2019-4 and FAPESP grant 2019/21281-4.
HN would also like to thank the ICTP-SAIFR for their support through FAPESP grant 2021/14335-0.
The work of PM is supported by FAPESP grant 2022/12401-9.
%
%
%%REFS

\bibliography{MomentInertia.bib}
\bibliographystyle{utphys}

%%%%%%%%%%%%%%%%%%%%%%%%%
\end{document}